\documentclass[pre,aps,showpacs,twocolumn,supergroupedaddress,epsfig,amsmath,amssymb,eqsecnum,floatfix]{revtex4}
\usepackage{epsfig,amsmath,amssymb,bm,epsf,graphics,psfrag,verbatim,subfigure,framed}
\usepackage[usenames,dvipsnames]{color}
\usepackage{bbm}
\usepackage{mathrsfs}
\usepackage{amsfonts}
\usepackage{color}
\usepackage{pbox}
\def\newblock{\hskip .11em plus .33em minus .07em}

\def\tv{\widetilde{v}}

\def\cosech{{\rm{cosech}}}

\def\xp{x_\perp}

\def\tvv{\widetilde{\mathbf{v}}}
\def\tv{\widetilde{v}}

\newcommand{\be}{\begin{equation}}
\newcommand{\ee}{\end{equation}}
\newcommand{\ba}{\begin{eqnarray}}
\newcommand{\ea}{\end{eqnarray}}
\newcommand{\bw}{\begin{widetext}}
\newcommand{\ew}{\end{widetext}}
\newcommand{\Tr}{{\rm{Tr}\,}}

\newcommand{\vv}{{\mathbf{v}}}

\newcommand{\Qv}{{\mathbf{Q}}}

\newcommand{\xv}{{\mathbf{x}}}

\newcommand{\uv}{{\bm{u}}}

\begin{document}

\title{Pseudo-Casimir stresses and elasticity of a confined elastomer film}
\author{Bing-Sui Lu$^1$}
\email{binghermes@gmail.com}
\author{Ali Naji$^2$}
\author{Rudolf Podgornik$^1$}
\affiliation{$^{1}$Department of Theoretical Physics, J. Stefan Institute, 1000 Ljubljana, Slovenia.\\
 $^{2}$School of Physics, Institute for Research in Fundamental Sciences (IPM), P.O. Box 19395-5531, Tehran, Iran
}

\date{\today}


\begin{abstract}
Investigations of the elastic behavior of bulk elastomers have traditionally proceeded on the basis of classical rubber elasticity, which regards chains as thermally fluctuating but disregards the thermal fluctuations of the cross-links. Here, we consider an incompressible and flat elastomer film of an axisymmetric shape confined between two large hard co-planar substrates, with the axis of the film perpendicular to the substrates. We address the impact that thermal fluctuations of the cross-links have on the free energy of elastic deformation of the system, subject to the requirement that the fluctuating elastomer cannot detach from the substrates. We examine the behavior of the deformation free energy for one case where a rigid pinning boundary condition is applied to a class of elastic fluctuations at the confining surfaces, and another case where the same elastic fluctuations are subjected to soft ``gluing" potentials. We find that there can be significant departures (both quantitative and qualitative) from the prediction of classical rubber elasticity theory when elastic fluctuations are included. Finally, we compare the character of the attractive part of the elastic fluctuation-induced, or pseudo-Casimir, stress with the standard thermal Casimir stress in confined but non-elastomeric systems, finding the same power law decay behavior when a rigid pinning boundary condition is applied; for the case of the gluing potential, we find that the leading order correction to the attractive part of the fluctuation stress decays inversely with the inter-substrate separation.

\end{abstract}

\maketitle

\section{Introduction}

There has been increasing interest in polymer network films of micron to nanoscale thicknesses~\cite{zhai_mckenna}, which can be used as sensors that mimic biological organs, tunable Bragg reflectors and synthetic cell substrata~\cite{Harsanyi, Zhai}. Natural polymer network films also exist, for example, the actin filament networks of cells and the intertwining sphingomyelin chains in a myelin sheath \cite{Terentjev}. A conventional picture of a polymer network is that of a collection of chains whose end-points (or junctures) are permanently cross-linked into an elastic matrix that undergoes deformation \cite{McKintosh}. The shear modulus emerges from the entropic cost of thermal fluctuations of the chains~\cite{treloar,james_guth,rubinstein_panyukov,wt}. On the other hand, the elastic matrix itself is also a thermally fluctuating entity at room temperature, with the chain end-points undergoing thermally driven displacements about their mean positions~\cite{james_guth,goldbart_vulcan,xgr,mao_long}. The end-points have a typical \emph{localization length} or root-mean-square displacement that is directly related to how densely the network is cross-linked~\cite{goldbart_vulcan}. Physically the typical localization length reflects how strongly localized the network constituents are, i.e., a smaller value would correspond to a stronger localization. We can also regard it as a cut-off length scale below which continuum elasticity theory no longer applies, and it is in this sense that we shall apply the term in the rest of the Paper. 
Thus quantities such as strain and deformation tensor are really coarse-grained quantities that have a meaning only on length scales larger than the typical localization length. The classical theory of rubber elasticity~\cite{treloar} (also known as the affine network model~\cite{rubinstein_panyukov}) takes into account the thermal fluctuations of the chains, but  regards the elastic matrix (i.e., the cross-linked ends of the fluctuating chains) as thermally \emph{non}-fluctuating. However, the effects of thermal fluctuations of the elastic matrix (henceforth called ``elastic fluctuations") can substantially modify the predictions of classical rubber elasticity when the local incompressibility of the elastomer is taken into account, as has been shown for the case of a \emph{bulk} elastomer (i.e., whose dimensions are all much larger than the typical localization length-scale)~\cite{xgr}. It is thus of interest to study how elastic fluctuations can modify the elastic deformation behavior of a \emph{thin} locally incompressible elastomer film confined between two large hard co-planar substrates. This would involve the interplay between elastic fluctuations and the finite-size effects of the elastomer system. Here and in what follows, we take ``thin" to refer to a thickness 
that is much smaller than the square root of the cross-sectional area of the elastomer surface co-planar with the substrates, but still larger than the typical localization length. 

Our paper represents a first step in the theoretical investigation of the effects of thermal elastic fluctuations on the behavior of a \emph{confined} incompressible elastomer film. In this confined system boundary conditions have to be imposed on the elastic fluctuations at the confining surfaces. We address the case of a boundary condition (BC) that enforces the vanishing of elastic fluctuations at the confining surfaces (i.e., the ``rigid pinning" BC, which is mathematically equivalent to a \emph{Dirichlet} BC), as well as the case where we have a ``soft" \emph{gluing} potential (instead of a ``hard" Dirichlet BC) acting at each confining surface on a class of elastic fluctuations. From studies in other systems, confinement has been known to result in fluctuation-induced, {\em Casimir} or \emph{pseudo-Casimir} stresses~\cite{bordag,mostepanenko,ajdari,ziherl,jure,kardar_review}. Such stresses are the result of  thermal or quantum fluctuations of a field with long range correlation, constrained by the presence of boundary surfaces. The surfaces ``feel" the presence of each other via these fluctuating fields. The long range correlations can emerge for example in ordered soft matter systems which exhibit broken symmetry (i.e., an ordered phase), where the corresponding Goldstone modes mediate the pseudo-Casimir force between the boundary surfaces of the system. In fluctuating elastomers, there are elastic phonons which are the coarse-grainings of thermally excited random displacements of the cross-linking points. As we will see later, these phonons are ``massless" (i.e., they only appear as gradient terms in the Hamiltonian), and it is well-known that the phonons can therefore exhibit long-range correlations~\cite{xgr}. Thus, we expect analogous pseudo-Casimir stresses to arise in a thermally fluctuating confined elastomer. In confined nematics~\cite{ajdari,ziherl} and confined polymer liquid crystals~\cite{jure}, the fluctuation-induced pseudo-Casimir stress is predicted to be attractive and decay as the third power of inverse inter-surface separation. The route is thus open to the following inter-related pair of investigations. Firstly, from the angle of rubber elasticity, how do thermal elastic fluctuations modify the elastic behavior predicted by classical rubber theory, for a confined elastomer film? In particular, how does the type of boundary conditions impact on the deformation free energy behavior? Secondly, from the purview of the field of fluctuation-induced forces~\cite{kardar_review}, how does the character of the \emph{attractive} part of the elastic fluctuation-induced stress in a confined elastomer compare with the Casimir stress induced by confined electromagnetic, nematic, or polymer liquid crystal fluctuation modes? 

In the interest of simplicity, we disregard the effects of disorder introduced by the heterogeneous distribution of cross-links, and we focus instead on \emph{homogeneous} elastomers, in which every point of the elastomer (together with its thermal fluctuations) undergoes an affine transformation under an externally applied \emph{uniaxial} shear deformation. We study the case of incompressible elastomers as the shear modulus of rubber is substantially smaller than its bulk modulus (often by at least two orders of magnitude~\cite{wt}), which justifies the approximation of incompressibility. We also limit our consideration to elastomer films whose thicknesses are larger than the typical localization lengthscale (and thus of \emph{macroscopic} lengthscales), in order that the framework of continuum elasticity theory may still be used, but the thickness is much smaller than the transverse dimension of the film, so that the finite-size effects on elastic fluctuations cannot be neglected. 

\section{The Hamiltonian}

Our system is an elastomer film which in its undeformed state has a certain thickness $L$ and cross-sectional area $S$, and is confined between a pair of co-planar hard substrates. 
The elastomer film is then subjected to a uniaxial shear deformation. 
To describe the corresponding energy cost for a given thermal configuration of the elastomer, we take the Hamiltonian of Ref.~\cite{xgr}:
\begin{equation}
H_{el} = \frac{\mu_0}{2}\int_0^{L} \!\! dz \!\int\!\!d^2\xp \frac{\partial R_i(\mathbf{x})}{\partial x_a}\frac{\partial R_i(\mathbf{x})}{\partial x_a},
\label{eq:H_el}
\end{equation}
where $\mathbf{R}(\mathbf{x}) = \underline{\underline{\Lambda}}\!\cdot\!(\mathbf{x}+\mathbf{u}(\mathbf{x}))$. Here $\underline{\underline{\Lambda}}$ is the deformation gradient, $\mathbf{x}$ is the mean position of a given mass-point in the elastomer prior to deformation, and $\mathbf{u}(\mathbf{x})$ denotes the elastic phonon, i.e., the thermal fluctuation of the mass-point about its mean position. The coordinate $\xv$ can equally well be regarded as a label affixed to each mass-point in the undeformed elastomer, which tags along as the mass-point moves when the elastomer is deformed. 

\section{Partition function}

The partition function is given by      
\begin{equation}
Z = \int\!\mathcal{D}\mathbf{u} \, 
\delta(\nabla\!\cdot\!\mathbf{u}) e^{-\beta H_{el} } 
\equiv Z_0 Z_\mathbf{u}.   
\label{eq:Z}
\end{equation}
Here we have decomposed $Z$ into a fluctuation-free, \emph{mean-field} part $Z_0$, and a fluctuation contribution $Z_{\mathbf{u}}$. The symbol $\delta(f)$ denotes the \emph{Dirac delta-functional}, which is defined to be zero (non-zero) if $f\neq 0$ ($f= 0$)~\cite{courant-hilbert}. By writing the Dirac delta-function inside the functional integral over $\mathbf{u}$, we are enforcing the local incompressibility of the elastomer: $\nabla\!\cdot\!\mathbf{u}=0$. This \emph{linear} constraint is true only for \emph{small} $\mathbf{u}$, which is the regime we consider. This constraint is derived from the more general \emph{nonlinear} local incompressibility constraint, viz., $\det \partial \mathbf{R}/\partial \mathbf{x} = 1$. We can see this by writing $\det \partial \mathbf{R}/\partial \mathbf{x} =\det (\underline{\underline{\Lambda}}) \det (\underline{\underline{\delta}} + \partial \mathbf{u}/\partial \mathbf{x})$, and using the global incompressibility of the elastomer, viz., $\det (\underline{\underline{\Lambda}})=1$, which forces $0 = \ln \det (\underline{\underline{\delta}} + \partial \mathbf{u}/\partial \mathbf{x}) = \rm{Tr}\, \ln (\underline{\underline{\delta}} + \partial \mathbf{u}/\partial \mathbf{x})$. For small $\mathbf{u}$ we can expand the logarithm to linear order, and obtain $\nabla\cdot\mathbf{u} = 0$. 

The mean-field free energy is given by
\begin{equation}
F_0 = -k_{\rm B}  T \ln Z_0   = \frac{\mu_0}{2}~V~\rm{Tr}\,(\underline{\underline{\Lambda}}^{{\rm T}} \cdot \underline{\underline{\Lambda}}).
\label{eq:Z0}
\end{equation}
Taking the $z$-direction to be perpendicular to the cross-sectional surface of the elastomer, a uniaxial shear deformation is described by the deformation gradient: $\underline{\underline{\Lambda}} = {\rm diag}(\lambda^{-1/2}, \lambda^{-1/2}, \lambda)$, where $\lambda > 1$ ($\lambda < 1$) corresponds to uniaxial extension (compression). Correspondingly, the mean-field free energy becomes
\begin{equation}
F_0 = \frac{\mu_0}{2} V \left( \lambda^2 + \frac{2}{\lambda} \right). 
\label{eq:F_0}
\end{equation}
The elastic fluctuation correction to $Z$ is described by
\begin{equation}
Z_\mathbf{u} \equiv 
\int\!\mathcal{D}\mathbf{u} \, \delta(\nabla\!\cdot\!\mathbf{u}) \,
e^{-\beta H_{\mathbf{u}} } 
\label{eq:Z_u}
\end{equation}
where $H_{\mathbf{u}}$ is the Hamiltonian for elastic fluctuations. 
As we show in Appendix~\ref{appa}, $H_{\mathbf{u}}$ is given by 
\begin{equation}
\label{eq:H_u}
H_{\mathbf{u}} = \frac{\mu_0}{2} \! \int \! d^3 x \, \partial_a u_b \Lambda_{bi}^{{\rm T}} \Lambda_{ic}^{~} \partial_a u_c.
\end{equation}
We can thus also express $H_{el}$ as
\begin{equation}
\label{eq:neweq}
H_{el} = \frac{\mu_0}{2} \int_0^{L} \!\! dz \!\int\!\!d^2\xp (\Lambda_{ai}^{{\rm T}} \Lambda_{ia}^{~} + \partial_a u_b \Lambda_{bi}^{{\rm T}} \Lambda_{ic}^{~} \partial_a u_c).
\end{equation}
The first term describes classical rubber elasticity, and is derived by considering the entropy of fluctuating Gaussian chains with end-points fixed in a thermally non-fluctuating matrix that deforms affinely. On the other hand, the second term allows for the thermal fluctuations of the end-points themselves. 

\subsection{Boundary conditions}
\label{sec:bc}

Furthermore, we need to specify boundary conditions (BC) for the elastic fluctuations $\uv$ at the two interfaces. Owing to the local incompressibility constraint, the BC can only be enforced on two components of $\uv$. 
Let us write $\uv = (\vv, u_z)$ and make a Helmholtz decomposition of $\vv$ into an irrotational and a solenoidal part: $\vv=\vv^{||} + \vv^\perp$. The \emph{solenoidal} fluctuation, $\vv^\perp$, satisfies $\nabla_\perp \cdot \vv^{\perp}=0$, whilst the \emph{irrotational} fluctuation, $\vv^{||}$, satisfies $\nabla_\perp\times\vv^{||}=0$. The symbol $\nabla_\perp \equiv (\partial_x, \partial_y)$ refers to the two-dimensional gradient operator. 
The first set of boundary conditions are the \emph{non-detachment} BC: 
\be
\label{eq:nondetachment_BC}
u_z(z=0)=u_z(z=L)=0, 
\ee
which enforce the condition that the surfaces of the elastomer do not detach from the substrates~\cite{raphael}. 
Regarding the second set of boundary conditions for the other components of $\uv$, we can have different choices depending on the physical make-up of the interfaces. For example, if the elastomer surfaces are rigidly pinned to the substrates so that the elastic displacements at the interfaces cannot undergo solenoidal motion, we can specify the \emph{rigid pinning} boundary condition for $\vv^\perp$, i.e., 
\be
\label{eq:pinned_BC}
\vv^\perp(z=0)=\vv^\perp(z=L)=0,
\ee
whilst the corresponding BC for $\vv^{||}$ can be found from the local incompressibility constraint. 
On the other hand, if we allow for some solenoidal ``slippage" of the elastomer film at the interfaces, then instead of the rigid pinning BC we have additional terms (the soft ``gluing" potentials, to be described in Sec.~\ref{sec:glue}) in $H_{el}$ that describe the energetic cost of slippage. In this case, the non-detachment BC still holds for $u_z$. 

We need to calculate the fluctuation correction to the mean-field behavior described by Eq.~(\ref{eq:F_0}). The fluctuation calculation is made somewhat more challenging by the presence of the local incompressibility constraint in the partition function. The corresponding strategy we adopt is to make use of a certain mode decomposition that automatically enforces the local incompressibility constraint and also shows the partition function to be a functional integral over two independent field degrees of freedom. We therefore next turn to the mode decomposition. 

\subsection{Mode decomposition}

Let us write (in real space) $\mathbf{u} = (\mathbf{v}, \phi)$ where $\mathbf{v} \equiv (u_x, u_y)$ and $\phi \equiv u_z$. Let us also define the differential operator in the $x-y$ directions: $\nabla_\perp \equiv (\partial_x, \partial_y)$. 
The uniaxially compressed elastomer is described by the Hamiltonian
\begin{eqnarray}
H_{\mathbf{u}} &\!\!=\!\!& \frac{\mu_0}{2} \! \int_0^{L} \!\! dz \! \int\!d^2\rho \left( \frac{1}{\lambda} \partial_i v_\mu \, \partial_i v_\mu + \lambda^2 \partial_i \phi \, \partial_i \phi \right) 
\nonumber\\
&\!\!=\!\!&  
\frac{\mu_0}{2} \! \int_0^{L} \!\! dz \! \int\!d^2\rho 
\Big( \frac{1}{\lambda} (\nabla_\perp \!\cdot\! \mathbf{v}^{||})^2 + \frac{1}{\lambda} (\partial_z\mathbf{v}^{||})^2 
\nonumber\\
&&
+ \frac{1}{\lambda} (\nabla_\perp \! \times \! \mathbf{v}^{\perp})^2 + \frac{1}{\lambda} (\partial_z\mathbf{v}^{\perp})^2 
+ \lambda^2 \partial_i \phi \, \partial_i \phi 
\Big) 
\label{eq:Hami}
\end{eqnarray}
where the Greek index $\mu=1,2$ and the Latin index $i=1,2,3$. 
As the co-planar substrates break translation symmetry in the $z$-direction but leave the system translationally invariant in the $x-y$ plane, the two-dimensional inverse Fourier transforms of $\mathbf{v}^{||}$ and $\mathbf{v}^{\perp}$ are given by 
\begin{eqnarray}
\mathbf{v}^{||}(\mathbf{\rho}, z) &=& \int \frac{d^2\Qv}{(2\pi)^2} e^{i\mathbf{\rho}\cdot\mathbf{Q}} \widetilde{\vv}^{||}({\mathbf{Q}}, z), 
\nonumber\\
\mathbf{v}^{\perp}(\mathbf{\rho}, z) &=& \int \frac{d^2\Qv}{(2\pi)^2} e^{i\mathbf{\rho}\cdot\mathbf{Q}} \tvv^{\perp}({\mathbf{Q}}, z), 
\label{eq:FT_v}
\end{eqnarray}
and the inverse Fourier transform of $\phi$ is given by 
\be
\label{eq:phi-FT}
\phi(\mathbf{\rho}, z) = \int \frac{d^2\Qv}{(2\pi)^2} e^{i\mathbf{\rho}\cdot\mathbf{Q}} \widetilde{\phi}({\mathbf{Q}}, z).  
\ee
Here $\Qv=(Q_x,Q_y)$ is the two-dimensional wave-vector conjugate to $\mathbf{\rho}=(x,y)$. 
In two-dimensional Fourier space the properties of $\vv^{||}$ and $\vv^{\perp}$ are described by $\widehat{Q}_\mu \tv_\mu^{\perp} = 0$ and $\epsilon_{\mu\nu}\widehat{Q}_\mu \tv_\nu^{||} = 0$ (where $\widehat{Q} \equiv \Qv/Q$, $\epsilon_{12}=-\epsilon_{21}=1$ and $\epsilon_{11}=\epsilon_{22}=0$), which implies they can be expressed in terms of scalar modes $\chi$ and $\psi$, viz., 
\begin{equation}
\label{eq:chi_psi}
\tv_\mu^{||} \equiv \widehat{Q}_\mu \chi, \quad {\tv}_\mu^{\perp} \equiv \epsilon_{\mu\nu} \widehat{Q}_\nu \psi.
\end{equation}
In terms of $\chi$, $\psi$ and $\widetilde{\phi}$, we can rewrite $H_{\mathbf{u}}$ as 
\ba
&&H_{\mathbf{u}}[\chi,\psi,\phi]
\nonumber\\ 
&=& 
\frac{\mu_0}{2\lambda} \! \int_0^{L} \!\! dz \! \int\!\frac{d^2\Qv}{(2\pi)^2} 
\Big(
Q^2 |\psi(\Qv,z)|^2 + |\partial_z \psi(\Qv,z)|^2
\nonumber\\
&&+
Q^2 |\chi(\Qv,z)|^2 + |\partial_z \chi(\Qv,z)|^2
\nonumber\\
&&+\lambda^3 (Q^2 |\widetilde{\phi}(\Qv,z)|^2 + |\partial_z \widetilde{\phi}(\Qv,z)|^2)
\Big)
\label{eq:Hamj}
\ea
The corresponding partition function is given by 
\begin{eqnarray}
Z_{\mathbf{u}} 
&\!\!=\!\!& 
\!\! 
\int\!\!\mathcal{D}\mathbf{\chi} \!\! 
\int\!\!\mathcal{D}\mathbf{\psi} \!\! 
\int\!\!\mathcal{D}\widetilde{\phi} \, 
\prod_{\{\Qv\}} \prod_{\{z\}}
\delta(Q\chi(\Qv,z)-i\partial_z\widetilde{\phi}(\Qv,z)) 
\nonumber\\
&&\times 
e^{-\beta H_{\mathbf{u}}[\chi,\psi, \widetilde{\phi}]}   
\label{eq:Zconstraint}
\end{eqnarray}
We turn next to the computation of the free energy for the two following types of boundary conditions for $\psi$: (i)~rigid pinning BC and (ii)~``gluing" potential (to be described in Sec.~\ref{sec:glue}). 

\section{Elastomer rigidly pinned at the interfaces}
\label{sec:rigidpinning}
\subsection{Fluctuation modes}
First we consider the case of an elastomer rigidly pinned at the interfaces with the co-planar substrates. This means that we implement both the non-detachment BC (Eq.~(\ref{eq:nondetachment_BC})) for $\widetilde{\phi}$ and the rigid pinning BC (Eq.~(\ref{eq:pinned_BC})) for $\psi$, so these fluctuation fields are given by sinusoidal Fourier series:
\begin{subequations}
\label{eq:fluct-series}
\be
\label{eq:phi-series}
\widetilde{\phi}(\Qv,z) = \sum_{n=1}^{\infty}\sqrt{\frac{2}{L}}\sin\left( \frac{n\pi z}{L} \right) (\phi_n^{\rm{re}}(\Qv)+i \phi_n^{\rm{im}}(\Qv)),
\ee
\be
\psi(\Qv,z) = \sum_{n=1}^{\infty}\sqrt{\frac{2}{L}}\sin\left( \frac{n\pi z}{L} \right) (\psi_n^{\rm{re}}(\Qv)+i \psi_n^{\rm{im}}(\Qv)).
\ee
\end{subequations}
Here the superscripts ``re" and ``im" refer to real and imaginary parts.  
The independent fluctuation degrees of freedom along the $z$-direction are now replaced by the independent discrete modes labeled by $n$, where $n=1,2,3,\dots$. 
To determine $\chi(\Qv,z)$, we return to the local incompressibility constraint: $-\partial_z ({\rm{Im}} \, \widetilde{\phi}(\Qv,z)) = Q \, {\rm{Re}} \, \chi(\Qv,z)$ and $\partial_z ({\rm{Re}} \, \widetilde{\phi}(\Qv,z)) = Q \, {\rm{Im}} \, \chi(\Qv,z)$. Used in conjunction with Eq.~(\ref{eq:phi-series}), we obtain 
\be
\label{eq:chi-series}
\chi(\Qv,z) = \sum_{n=1}^{\infty}\sqrt{\frac{2}{L}}\left( \frac{n\pi}{QL} \right)\cos\left( \frac{n\pi z}{L} \right) (-\phi_n^{{\rm im}} + i\phi_n^{{\rm re}}).
\ee
In terms of the discrete modes we can write Eq.~(\ref{eq:Hamj}) as 
\ba
H_{\mathbf{u}}[\{\psi_n, \phi_n\}] &\!\!=\!\!& \frac{\mu_0}{2\lambda}\sum_{n=1}^{\infty} \int\!\! \frac{d^2\Qv}{(2\pi)^2} 
\Big\{
\Big( Q^2 + \left( \frac{n\pi}{L} \right)^2 \Big) 
\nonumber\\
&&\quad\times 
\left[ (\psi_n^{{\rm re}}(\Qv))^2 + (\psi_n^{{\rm im}}(\Qv))^2 \right] 
\nonumber\\
&&+\Big( \lambda^3 Q^2 + \Big(\frac{n\pi}{L}\Big)^2 \Big)\Big( 1 + \Big( \frac{n\pi}{QL} \Big)^2 \Big) 
\nonumber\\
&&\quad\times 
\left[ (\phi_n^{{\rm re}}(\Qv))^2 + (\phi_n^{{\rm im}}(\Qv))^2 \right] 
\Big\}
\label{eq:Hfluctcon}
\ea
As we show in Appendix~\ref{appb}, this leads to the following fluctuation contribution to the free energy of elastic deformation: 
\be
F_{\mathbf{u}}(\lambda) = 
\frac{k_{{\rm B}}T S}{2} \! 
\sum_{n=1}^{\infty} \! \int \!\! \frac{d^2\Qv}{(2\pi)^2}
\ln \Big( \lambda Q^2 + \frac{1}{\lambda^2}\Big( \frac{n\pi}{L} \Big)^2 \Big)
\label{eq:Fu_discrete}
\ee
We can rewrite Eq.~(\ref{eq:Fu_discrete}) as 
\be
F_{\mathbf{u}}(\lambda) = \frac{k_{{\rm B}}TS}{4\pi}\int_0^{\pi\xi^{-1}} \!\!\!\!\!\!\!\! dQ\,Q \sum_{n=1}^{M} \big[ \ln(a^2+n^2) -2 \ln \lambda + B \big], 
\label{eq:Fu_lamb}
\ee
where $a \equiv \lambda^{3/2} QL/\pi$, $B \equiv \pi^2 M k_{{\rm B}}TS / 4 L \xi^2$ is a term independent of $\lambda$ (and which we will thus ignore), 
and we have set an upper limit $M \equiv L/\xi$ on the discrete sum, as the number of fluctuating degrees of freedom in the $z$-direction is limited by the typical localization length $\xi$. 

\subsection{Fluctuation free energy}
Before we turn to evaluate the free energy of the confined elastomer film, we make a few general remarks about the anticipated features of such a free energy. In a confined elastomer film, thermal fluctuations of the elastic matrix introduce qualitative changes to the free energy and the value of $\lambda$ that minimizes the free energy. In an isotropic bulk elastomer, such elastic fluctuations cause a spontaneous change in the volume of the undeformed system (relative to the state of the undeformed elastomer in mean-field theory~\cite{mao_long}), but the value of $\lambda$ that characterizes the undeformed state remains unchanged at unity~\cite{xgr}. The latter is expected on grounds of symmetry as it is equally energetically costly for elastic fluctuations to occur in every direction in space, the elastomer being equally macroscopically large (and homogeneous) in every direction. In this respect, the reference (i.e., undeformed) space can be regarded as isotropic and translationally invariant in every direction. On the other hand, the reference space of a confined elastomer film is isotropic only in the transverse directions, and translation symmetry is broken in the direction normal to the film's surface. As we have seen, applying the Dirichlet-type non-detachment BC on the spectral decomposition of the elastic fluctuations leads to a \emph{discrete} spectrum of modes in the normal direction, whereas there is a \emph{quasi-continuous} spectrum of modes in the transverse directions. Each mode carries a thermal energy, and there are many more modes in the transverse directions than in the normal direction. The \emph{spectral} anisotropy will thus be reflected in the \emph{energetic} anisotropy of the fluctuation-corrected free energy, which means for example that we expect that the free energy minimum should occur at a value of $\lambda$ different from unity, $\lambda$ being the strain measured relative to the \emph{isotropic} undeformed state (or ground state) in mean-field theory. Hence, the fluctuations generate internal ``pre-stress" that causes the system to undergo a spontaneous shear relaxation, while the physically measurable strain is defined with reference to the state that has already spontaneously relaxed. 
\begin{figure}
\begin{center}
\includegraphics[width=0.48\textwidth]{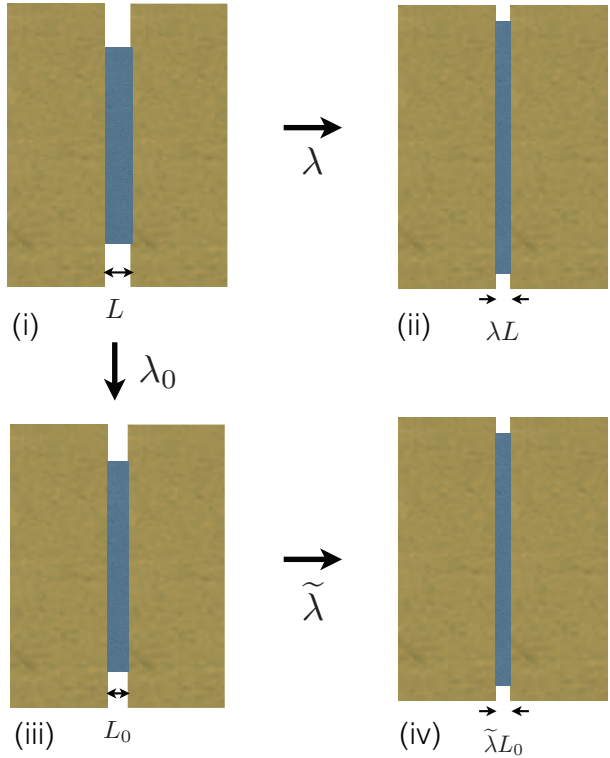}
\end{center}
\caption{Comparison of different ground states and relative deformations. The state in (i) corresponds to the undeformed or reference state in mean-field theory (where thermal fluctuations of the elastic matrix are absent). We call this the \emph{mean-field} ground state, and is the ground state considered by classical rubber elasticity theory. In (ii), the elastomer undergoes an external deformation with a deformation $\lambda$ measured relative to the mean-field ground state. If we allow for the effect of thermal fluctuations of the elastic matrix, the elastomer in state (i) will undergo a spontaneous relaxation to state (iii), with a deformation $\lambda_0$ measured relative to the mean-field ground state. We call state (iii) the fluctuation corrected or \emph{true} ground state. In (iv) the elastomer undergoes an external deformation $\widetilde{\lambda}$ relative to the true ground state, and such deformation corresponds to what is measured in experiment.} 
\label{fig:ground_states}
\end{figure}

To evaluate the free energy, we first decompose the discrete sum in Eq.~(\ref{eq:Fu_lamb}) as follows: 
\ba
\sum_{n=1}^{M} \ln(n^2+a^2) &=& \sum_{n=1}^{M} (\ln(n+ia) + \ln(n-ia))
\nonumber\\
&=& \ln((1+ia)_M + (1-ia)_M)
\nonumber\\
&=& \ln\left[ \frac{\Gamma(M+1-ia)\Gamma(M+1+ia)}{\Gamma(1+ia)\Gamma(1-ia)} \right]
\nonumber
\ea
In the above, the Pochhammer symbol $(x)_n$ denotes $\Gamma(x+n)/\Gamma(x)$. We make use of the result (see e.g., Ref.~\cite{ww}) 
\be
\Gamma(1+ia)\Gamma(1-ia) = \frac{\pi a}{\sinh(\pi a)}
\label{eq:Gamma1}
\ee
and in the limit that $z \gg 1$, use Stirling's approximation to $\Gamma(z)$~\cite{ww}
\be
\Gamma(z) \approx (z/e)^z\sqrt{2\pi/z}.
\ee
We thus have 
\ba
&&\Gamma(M+1-ia)\Gamma(M+1+ia) 
\nonumber\\
&\approx& 
2\pi \, e^{(M+1-ia)\ln(M+1-ia) - (M+1-ia)} 
\nonumber\\
&&\times 
e^{(M+1+ia)\ln(M+1+ia) - (M+1+ia)}
\nonumber\\
&& \times 
e^{-\frac{1}{2}\ln(M+1-ia)-\frac{1}{2}\ln(M+1+ia)}
\nonumber\\
&=&
2\pi \, e^{(M+\frac{1}{2})\ln((M+1)^2+a^2)} 
\nonumber\\
&&\times 
e^{-2a\tan^{-1}(\frac{a}{M+1}) - 2(M+1)},
\ea
where we have used the identity $\tan^{-1} z = (i/2) \ln((1-iz)/(1+iz))$~\cite{wolfram}. 
By writing $L/\xi = M \approx M+1/2$, we have
\ba
&&\ln ( \Gamma(M+1-ia)\Gamma(M+1+ia) )
\nonumber\\
&\approx& 
\ln 2\pi + \frac{L}{\xi}\ln \Big[ \Big( \frac{L}{\xi} \Big)^2 + \lambda^3 \Big( \frac{Q L}{\pi} \Big)^2 \Big]
\nonumber\\
&&
- \frac{2\lambda^{3/2} Q L}{\pi}\tan^{-1}\left( \frac{\lambda^{3/2}Q\xi}{\pi} \right) - \frac{2L}{\xi}.
\label{eq:GammaM}
\ea


\begin{figure}[t!]\begin{center}
	\begin{minipage}[b]{0.47\textwidth}\begin{center}
		\includegraphics[width=\textwidth]{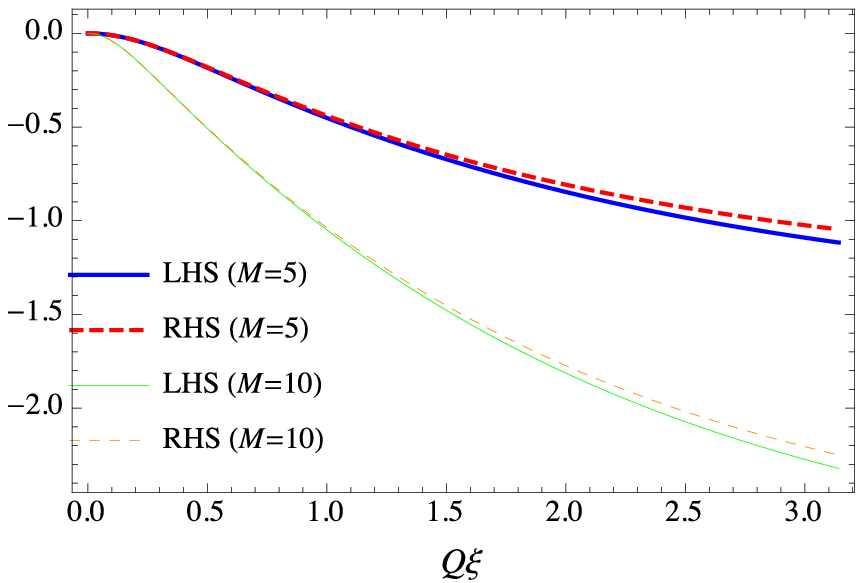} (i)
	\end{center}\end{minipage}\vskip.3cm
	\begin{minipage}[b]{0.47\textwidth}\begin{center}
		\includegraphics[width=\textwidth]{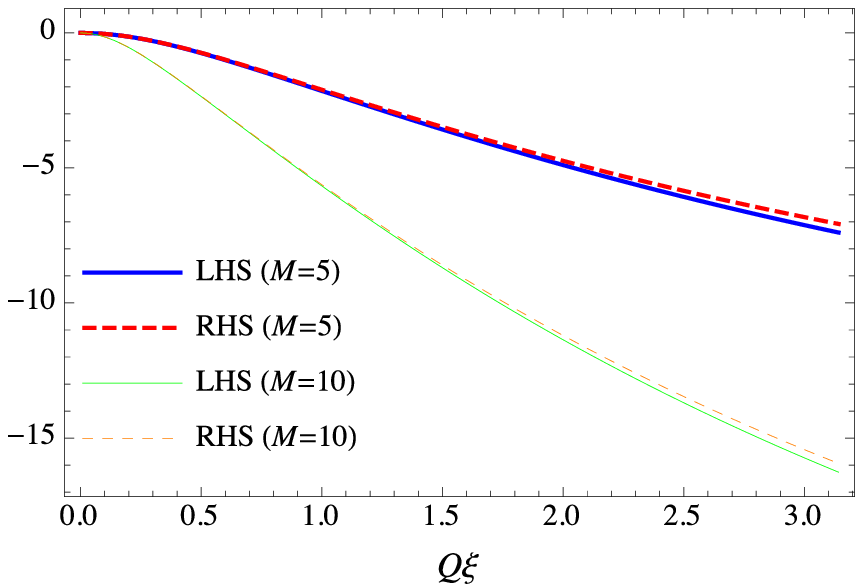} (ii)
	\end{center}\end{minipage} 	
\caption{Comparison of the left-hand side (LHS) and right-hand side (RHS) of Eq.~(\ref{eq:discrete_sum}), for (i)~$\lambda=0.9$ and (ii)~$\lambda=0.3$, with the LHS and RHS evaluated relative to their values at $\lambda=1$. For each case we plot the behavior for $M=5$ and $M=10$.}
\label{fig:lhsrhs}
\end{center}
\end{figure}

Using (\ref{eq:Fu_lamb}), (\ref{eq:Gamma1}) and (\ref{eq:GammaM}), the discrete sum can be put in the form
\ba
\label{eq:discrete_sum}
&&\sum_{n=1}^{M} \ln(a^2 + n^2) 
\\
&\approx& M \ln \left[ 1+ \left( \frac{a}{M} \right)^2 \right] + 2a\cot^{-1}\left( \frac{a}{M} \right) 
\nonumber\\
&&- \ln \pi a + \ln(1-e^{-2\pi a}) + C,
\nonumber
\ea
where we used the identity $\tan^{-1} (z) = \pi/2 - \cot^{-1} (z)$ and approximated $M+1$ and $M+\frac{1}{2}$ by $M$, which is valid for large $M$. 
The term $C \equiv \ln \pi + 2M \ln M - 2M$ is independent of $\lambda$, and we can thus disregard the corresponding term in the free energy. 
In Fig.~\ref{fig:lhsrhs}, we compare the left-hand side (LHS) and right-hand side (RHS) of Eq.~(\ref{eq:discrete_sum}), with the LHS and RHS evaluated relative to their values at $\lambda=1$, for $M=5$ and $M=10$, and $\lambda=0.3$ and $\lambda=0.9$. We see that the agreement between the exact LHS and its approximation in the RHS improves for larger values of $M$ and smaller values of $\lambda$. Furthermore, for each value of $\lambda$ and $M$, the agreement is better for the lower half-range of values of $Q$, with the error becoming  more noticeable only for $Q$ close to the upper bound $\pi\xi^{-1}$ (the upper bound being there because continuum elasticity is not defined on lengthscales smaller than the typical localization length). 

Writing $M = L/\xi$, we have
\ba
\label{eq:Fu_comp}
&&F_{\mathbf{u}}(\lambda) 
\\
&\approx&  
\frac{k_{{\rm B}}TS}{4\pi} \!\!
\int_0^{\pi\xi^{-1}} \!\!\!\!\!\!\!\! dQ\,Q \Big\{
\ln(1-e^{-2\lambda^{3/2}QL})
-\ln \lambda^{3/2}
\nonumber\\
&&
- \frac{2L}{\xi}\ln\lambda + \frac{L}{\xi}\ln\Big[ 1 + \lambda^3 \Big( \frac{Q \xi}{\pi} \Big)^2 \Big]
\nonumber\\
&&+\frac{2\lambda^{3/2}QL}{\pi}\cot^{-1}\Big( \frac{\lambda^{3/2}Q\xi}{\pi} \Big) 
\Big\}
\nonumber
\ea 
where we have neglected terms independent of $\lambda$. The first term describes the interaction between the confining surfaces, the second term can be interpreted as a surface tension term, and the remaining terms are the bulk terms. The first term resembles, but is not identical to, the standard form of a Casimir free energy (see, e.g., Ref.~\cite{ajdari}), the differences being the finite upper cut-off and the presence of $\lambda$ in the exponent. Strictly speaking, even in studies of the Casimir effect, a finite upper cut-off (which corresponds to the smallest length scale in the problem) should be imposed~\cite{mostepanenko-trunov}; however, the difference between the integral with a finite upper cut-off and one with an infinite upper bound is an irrelevant constant in the usual Casimir problems, which do not involve time-persistent elastic stresses in the intervening medium, and can thus be ignored. In our present problem, we cannot replace the integral by one with an infinite upper bound, because the difference depends on $\lambda$ and thus changes as the elastomer is deformed. 

We can rewrite the integral over the first term as the sum 
\ba
\label{eq:finiteintegral}
&&\int_{0}^{\pi \xi^{-1}} \!\!\!\!\!\!\!\! dQ\,Q 
\ln(1-e^{-2\lambda^{3/2}QL})
\\
&=&
-\frac{\zeta_{\rm{R}}(3)}{4\lambda^3 L^2} - g(\lambda,L),
\nonumber\\
&&g(\lambda,L) \equiv \int_{\pi \xi^{-1}}^{\infty} \!\!\!\!\!\!\!\! dQ\,Q 
\ln(1-e^{-2\lambda^{3/2}QL}).
\nonumber
\ea
In the above, $\zeta_{{\rm R}}(s) \equiv \sum_{n=1}^\infty 1/n^s$ is the Riemann zeta function, and $\zeta_{{\rm R}}(3) \approx 1.202$~\cite{ww}.  We thus obtain for the fluctuation free energy: 
\ba
F_{\mathbf{u}}(\lambda) 
&\approx& 
-\frac{k_{{\rm B}}T S \zeta_{\rm{R}}(3)}{16\pi\lambda^3 L^2} 
- \frac{k_{\rm{B}} T S}{4 \pi} g(\lambda,L)
\nonumber\\
&&- 
\frac{3\pi k_{{\rm B}}TS}{16\xi^2}\ln\lambda
+ \frac{\pi k_{{\rm B}}TV}{24\xi^3} f_{{\rm bulk}}.
\label{eq:Fusb}
\ea
In the above, we denote the volume by $V \equiv SL$, and 
\ba
\label{eq:fbulk}
f_{{\rm bulk}} &\equiv& 
4\lambda^{3/2}\cot^{-1}(\lambda^{3/2}) - 6 \ln \lambda -1 
\nonumber\\
&&+ \Big( 3 + \frac{1}{\lambda^3} \Big) \ln (1+\lambda^3).
\ea
The first three terms in Eq.~(\ref{eq:Fusb}) scale as $S$, whereas the rest are bulk terms that scale as $V$~\cite{footnote:anisotropy}. 

\begin{figure}
\begin{center}
\includegraphics[width=0.47\textwidth]{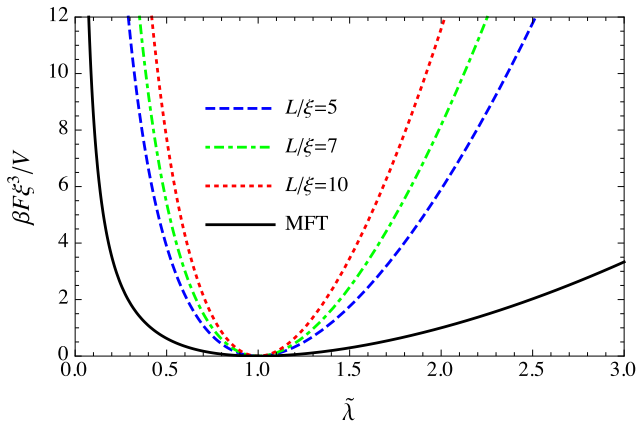}
\end{center}
\caption{Comparison of the free energy densities of elastic deformation as a function of deformation $\widetilde{\lambda}$ measured relative to the true ground state (i.e., undeformed state of the elastomer after elastic fluctuations have been allowed to relax), for $\mu_0 = k_{\rm{B}}T/\xi^3$ and (i)~$L=5\xi$ (blue, dashed), (ii)~$L=7\xi$ (green, dot-dashed), and (iii)~$L=10\xi$ (red, dotted). Comparison is made with the mean-field elastic energy in Eq.~(\ref{eq:F_0}) (black).} 
\label{fig:free_energy}
\end{figure}

\subsection{Reference state}
The full free energy is given by the sum of the mean-field contribution (Eq.~(\ref{eq:F_0})) and the fluctuation correction: 
\be
F_{{\rm{full}}} = F_0 + F_{\mathbf{u}}.
\ee
$F_0$ has a minimum at $\lambda=1$ but the minimum of $F_{{\rm{full}}}$ occurs at $\lambda = \lambda_0 \neq 1$. We can understand this by noting that $\lambda=1$ is the ground state of the mean-field theory, viz., 
\be
\left.\frac{\partial F_0}{\partial \lambda}\right\vert_{\lambda=1} = 0,
\ee
which corresponds to a thickness $L$ measured in a state where the positions of the cross-links (i.e., the ambient elastic matrix) do not undergo thermal fluctuations. On the other hand, in a state where the elastic matrix \emph{does} undergo thermal fluctuations, the fluctuations will cause the system to relax to a new equilibrium thickness $L_0$ distinct from $L$ (see Fig.~\ref{fig:ground_states}). The corresponding value of $\lambda$ (where $\lambda$ is a deformation relative to the \emph{mean-field} ground state) is $\lambda_0 \equiv L_0/L$, and $\lambda_0$ is determined from the stationarity condition: 
\be
\left.\frac{\partial F_{{\rm{full}}}}{\partial \lambda}\right\vert_{\lambda=\lambda_0} = 0. 
\label{eq:F_stationarity}
\ee
The value of $\lambda_0$ can be determined numerically. 
For example, for $\mu_0 = k_{\rm{B}}T/\xi^3$, $\lambda_0 \approx 1.0466$ for $L=5\xi$, $\lambda_0 \approx 1.03717$ for $L=7\xi$ and $\lambda_0 \approx 1.03008$ for $L=10\xi$. 
We call the state that satisfies Eq.~(\ref{eq:F_stationarity}), the \emph{true ground state}. If we measure a subsequent, isothermal, externally applied deformation relative to the \emph{true} ground state, the corresponding strain $\widetilde{\lambda}=L'/L_0$ (where $L'$ is the thickness of the deformed elastomer) is related to $\lambda$ via 
\be
\lambda = \widetilde{\lambda} \lambda_0. 
\ee  

\subsection{Deformation free energy}
The full deformation free energy is given by 
\be
\label{eq:deform_free_energy}
F(\widetilde{\lambda}) = F_{{\rm{full}}}(\widetilde{\lambda} \lambda_0) - F_{{\rm{full}}}(\lambda_0).
\ee
This quantity vanishes for zero external deformation relative to the true ground state [$F(\widetilde{\lambda}=1)=0$]. 
Using Eqs.~(\ref{eq:F_0}) and (\ref{eq:Fu_comp}), we find the deformation free energy for a system with rigid pinning BC:
\ba
&&(\beta \xi^3/V) F(\widetilde{\lambda})
\\
&=&
\frac{\beta \xi^3 \mu_0}{2} \left[ (\widetilde{\lambda}\lambda_0)^2+\frac{2}{\widetilde{\lambda}\lambda_0}-\lambda_0^2-\frac{2}{\lambda_0} \right] 
\nonumber\\
&&+ \frac{\xi^3}{4\pi L} \! \int_{0}^{\pi \xi^{-1}} \!\!\!\!\!\!\!\! dQ\,Q 
\ln\left[ \frac{1-e^{-2(\widetilde{\lambda}\lambda_0)^{3/2}QL}}{1-e^{-2\lambda_0^{3/2}QL}} \right]
\nonumber\\
&&+\frac{\pi}{6} \Big[ 
(\widetilde{\lambda}\lambda_0)^{3/2}\cot^{-1}((\widetilde{\lambda}\lambda_0)^{3/2}) 
-\lambda_0^{3/2}\cot^{-1}(\lambda_0^{3/2}) 
\Big]
\nonumber\\
&&+\frac{\pi}{24}\bigg[ 3 + \frac{1}{(\widetilde{\lambda}\lambda_0)^3} \bigg] \ln(1+(\widetilde{\lambda}\lambda_0)^3)
\nonumber\\
&&-\frac{\pi}{24}\bigg[ 3 + \frac{1}{\lambda_0^3} \bigg] \ln(1+\lambda_0^3) 
- \frac{\pi}{4}\ln \widetilde{\lambda}
- \frac{3\pi \xi}{16 L} \ln \widetilde{\lambda}.
\nonumber
\ea
In Fig.~\ref{fig:free_energy}, we show the behavior of the free energy density as a function of $\widetilde{\lambda}$ for three different thicknesses of the elastomer film, $L=5\xi$, $7\xi$ and $10\xi$~\cite{footnote:thicknesses}. We see that elastic fluctuations introduce a significant deviation from the qualitative behavior predicted by mean-field theory; in particular, fluctuations raise the overall free energy of the system relative to what mean-field theory predicts, and the increase is more significant for larger film thicknesses. For example, the full free energy can be seventy times larger than the mean-field value if an elastomer of undeformed thickness $5\xi$ is compressed to $70\%$ of its original thickness. 

\section{Elastomer ``glued" at the interfaces}
\label{sec:glue}

\subsection{Hamiltonian}
Next, we consider the case of an elastomer that is ``glued" at the interfaces, so that the solenoidal elastic fluctuations can undergo some slippage there. This is analogous, for example, to hydrodynamic slippage of a fluid in a nanopore due to a hydrophobic mismatch between the fluid and the surface of the nanopore~\cite{bocquet_charlaix_review}. In practice the glue can be an adhesive such as polysaccharide adhesive viscous exopolysaccharide (PAVE) isolated from the marine bacterium \emph{Alteromonas colwelliana}~\cite{smith-callow}. Instead of the rigid pinning BC we have a ``soft" \emph{gluing} potential at the confining surfaces~\cite{dean_softBC,schmidt,cavalcanti,bajnok,romeo}, described by two extra terms in $H_{u}$: 
\begin{eqnarray}
H_{\mathbf{u}} 
&=&  
\frac{\mu_0}{2} \! \int_0^{L} \!\! dz \! \int\!d^2\rho 
\Big( \frac{1}{\lambda} (\nabla_\perp \!\cdot\! \mathbf{v}^{||})^2 + \frac{1}{\lambda} (\partial_z\mathbf{v}^{||})^2 
\nonumber\\
&&
+ \frac{1}{\lambda} (\nabla_\perp \times \mathbf{v}^{\perp})^2 + \frac{1}{\lambda} (\partial_z\mathbf{v}^{\perp})^2 
+ \lambda^2 \partial_i \phi \, \partial_i \phi 
\Big) 
\nonumber\\
&&+ \frac{K_t}{2w_0^2} \! 
\int \! d^{2}\rho \left( 
|\vv^\perp(z=0)|^2
+
|\vv^\perp(z=L)|^2
\right)
\nonumber\\
\end{eqnarray}
The terms within the integral over $z$ are the same as those in Eq.~(\ref{eq:Hami}), whereas the last two terms describe the energetic cost of elastomer slippage at the interface. $K_t$ quantifies the slippage energy cost (or adhesion strength) per unit area on each of the two interfaces, and $w_0$ has the meaning of a ``slippage length". 
We have only written down soft BC terms for $\vv^\perp$ as the other component $\vv^{||}$ is completely determined by $\phi$ via the local incompressibility constraint. 
In terms of $\widetilde{\phi}$, $\chi$ and $\psi$ (defined by Eqs.~(\ref{eq:phi-FT}) and (\ref{eq:chi_psi})), we now have 
\ba
\label{eq:Hglue}
&&H_{\mathbf{u}}[\chi,\psi,\widetilde{\phi}]
\\ 
&=& 
\frac{\mu_0}{2\lambda} \! \int_0^{L} \!\! dz \! 
\int\!\frac{d^2\Qv}{(2\pi)^2} 
\Big(
Q^2 |\psi(\Qv,z)|^2 + |\partial_z \psi(\Qv,z)|^2
\nonumber\\
&&+
Q^2 |\chi(\Qv,z)|^2 + |\partial_z \chi(\Qv,z)|^2
\nonumber\\
&&+\lambda^3 (Q^2 |\widetilde{\phi}(\Qv,z)|^2 + |\partial_z \widetilde{\phi}(\Qv,z)|^2)
\Big)
\nonumber\\
&&+ \frac{K_t}{2w_0^2} \! 
\int \!\! \frac{d^2\Qv}{(2\pi)^2} \! 
\left( 
|\psi(\Qv,z=0)|^2 
+ 
|\psi(\Qv,z=L)|^2
\right).
\nonumber
\ea
The form of the above expression is consistent with the recovery of Eq.~(\ref{eq:fluct-series}) in the limit that $K_t \rightarrow \infty$ (which corresponds to the regime of rigid pinning). 

The non-detachment BC (Eq.~(\ref{eq:nondetachment_BC})) taken together with the local incompressibility condition means that $\phi$ and $\chi$ are still given by Eqs.~(\ref{eq:phi-series}) and (\ref{eq:chi-series}), respectively. The Hamiltonian is then given by
\be
H_{\mathbf{u}}[\{ \phi_n^{{\rm re}}, \phi_n^{{\rm im}} \}, \psi] = H_1 + H_\psi, 
\ee
where
\ba
&&H_1 \equiv \frac{\mu_0}{2\lambda} \sum_{n=1}^{\infty} \int\!\! \frac{d^2\Qv}{(2\pi)^2} 
\Big( \lambda^3 Q^2 + \Big(\frac{n\pi}{L}\Big)^2 \Big)\Big( 1 + \Big( \frac{n\pi}{QL} \Big)^2 \Big) 
\nonumber\\
&&\quad\qquad\times 
\left[ (\phi_n^{{\rm re}}(\Qv))^2 + (\phi_n^{{\rm im}}(\Qv))^2 \right]
\label{eq:H1_glue}
\ea
and
\ba
&&H_\psi \equiv    
\frac{\mu_0}{2\lambda} \int_{0}^{L} \!\! dz \! \int\!\! \frac{d^2\Qv}{(2\pi)^2} 
( Q^2 |\psi(\Qv,z)|^2 + |\partial_z \psi(\Qv,z)|^2 ) 
\nonumber\\
&&\quad\qquad+ 
\frac{K_t}{2w_0^2} 
\int\!\! \frac{d^2\Qv}{(2\pi)^2} 
\left( 
|\psi(\Qv,z=0)|^2
+
|\psi(\Qv,z=L)|^2
\right). 
\nonumber\\
\ea
Next, we turn to evaluate the partition function. 

\subsection{Partition function}

The fluctuation contribution to the partition function can be expressed as 
\be
\label{eq:Z_glue}
Z_{\mathbf{u}} = Z_1 Z_\psi
\ee
where 
\ba
Z_1 &\equiv&  
\prod_{\overset{\{ \Qv > \mathbf{0} \}}{\{ n \in Z^+ \}}} \!\!\!\! 
\int \! d\phi_{n}^{\rm{re}}(\Qv) \! \int \! d\phi_{n}^{\rm{im}}(\Qv) \, 
e^{-\beta H_1}
\label{eq:Z1_glue}
\\
Z_\psi &\equiv& \prod_{\overset{\{ \Qv > \mathbf{0} \}}{\{ z \in [0, L] \}}} \!\!\!\! 
\int \! d\psi^{\rm{re}}(\Qv, z) \! \int \! d\psi^{\rm{im}}(\Qv, z) \, 
e^{-\beta H_\psi}
\label{eq:Zpsi_glue}
\ea
In the above, $Z^+$ refers to the set of all positive integers and $\{ \Qv > \mathbf{0} \}$ refers to the set of all positive wave-vectors  (``positivity" being defined with reference to a straight line that divides the two-dimensional lattice of points $(Q_x, Q_y)$ into two halves; e.g., if we denote the normal vector to such a line by $\mathbf{n}$, then a wave-vector is positive if it satisfies $\Qv \cdot \mathbf{n} > 0$). 
To evaluate $Z_\psi$ we note that it has the form of a (Euclidean) Feynman path integral for a harmonic oscillator where $z$ is a time-like coordinate, and accordingly we apply the Fourier series method of Feynman and Hibbs~\cite{feynman-hibbs}. 
Let us define the Hamiltonian density $\widehat{H}_\psi(\Qv)$ in $\Qv$-space: 
\be
H_\psi \equiv \int\!\! \frac{d^2\Qv}{(2\pi)^2} \widehat{H}_\psi(\Qv),
\ee
and write 
\be
\psi(\Qv,z) = \psi_{{\rm cl}}(\Qv,z) + q(\Qv, z)
\ee
where $\psi_{{\rm cl}}(\Qv,z)$ is a solution to the saddle-point equation
\be
\partial_z^2 \psi_{{\rm cl}} = Q^2 \psi_{{\rm cl}}
\ee
and have the values $\psi_{{\rm cl}}(\Qv, z=0) = X_\Qv$ and $\psi_{{\rm cl}}(\Qv, z=L) = Y_\Qv$ at the boundary interfaces, whilst  
\be
\label{eq:qfluct}
q(\Qv, z) = \sum_{n} \psi_n(\Qv) \sin\left( \frac{n\pi z}{L} \right)
\ee
and $q(\Qv, z)$ satisfy Dirichlet boundary conditions. 
For fluctuations of given wave-vector $\Qv$ and specified boundary values $\psi(\Qv,z=0)=X_\Qv$ and $\psi(\Qv,z=L)=Y_\Qv$, we have made a decomposition into (i)~a ``classical trajectory" $\psi_{{\rm cl}}(\Qv,z)$ that extremizes the Boltzmann weight factor $e^{-\beta \widehat{H}_\psi(\Qv)}$, and (ii)~deviations $q(\Qv, z)$ about this trajectory, with the same ``end-points" (i.e., zero fluctuation amplitude at the boundaries). [Note that the case of $\psi(\Qv, z)$ subject to Dirichlet BC (considered in Sec.~\ref{sec:rigidpinning}) is a special case where $\psi_{{\rm cl}}(\Qv, z) \equiv 0$.] Tracing over all fluctuations in $Z_\psi$ is then equivalent to tracing over all deviations $q$ for the same classical trajectory for given boundary values, and then tracing over all possible boundary values. The above-mentioned decomposition also ensures that the Hamiltonian separates into two decoupled contributions:
\ba
\widehat{H}_\psi(\Qv) 
&=& 
\frac{\mu_0}{2\lambda} \int_{0}^{L} \!\! dz 
( Q^2 |\psi_{{\rm cl}}(\Qv,z)|^2 + |\partial_z \psi_{{\rm cl}}(\Qv,z)|^2 ) 
\nonumber\\
&&+ 
\frac{K_t}{2w_0^2} 
\left( 
|\psi_{{\rm cl}}(\Qv,z=0)|^2
+
|\psi_{{\rm cl}}(\Qv,z=L)|^2
\right) 
\nonumber\\
&&+ 
\frac{\mu_0}{2\lambda} \int_{0}^{L} \!\! dz 
( Q^2 |q(\Qv,z)|^2 + |\partial_z q(\Qv,z)|^2 ). 
\nonumber\\ 
\label{eq:deceq}
\ea
Accordingly, we find after implementing boundary conditions that
\be
\label{eq:psicl}
\psi_{{\rm cl}}(\Qv,z) = \frac{Y_\Qv - X_\Qv \cosh Q L}{\sinh Q L} \sinh Q z + X_\Qv \cosh Q z.
\ee
Substituting Eqs.~(\ref{eq:qfluct}) and (\ref{eq:psicl}) into Eq.~(\ref{eq:deceq}) and summing over wave-vectors, we obtain
\ba
&&H_\psi 
\equiv \sum_{n}  \int\!\! \frac{d^2\Qv}{(2\pi)^2} \frac{\mu_0 L}{4\lambda} \left( Q^2 + \frac{n^2\pi^2}{L^2} \right) 
\nonumber\\
&&\qquad\qquad\times
\left[ (\psi_{n}^{\rm{re}}(\Qv))^2 + (\psi_{n}^{\rm{im}}(\Qv))^2 \right]
\nonumber\\
&&+ \frac{1}{2}\int\!\! \frac{d^2\Qv}{(2\pi)^2} 
\bigg[
\left( \frac{\mu_0 Q}{\lambda}\coth Q L + \frac{K_t}{w_0^2} \right) (|X_\Qv|^2 + |Y_\Qv|^2)
\nonumber\\
&&\qquad-
\frac{2 \mu_0 Q}{\lambda} ( \cosech \, Q L )
( X_\Qv^{{\rm re}} Y_\Qv^{{\rm re}} + X_\Qv^{{\rm im}} Y_\Qv^{{\rm im}})
\bigg].
\label{eq:Hpsi_glue}
\ea
The evaluation of $Z_1$ and $Z_\psi$ in Eq.~(\ref{eq:Z_glue}) involves a straightforward Gaussian functional integration, and is carried out in Appendix~\ref{appc}. The result for $Z_{\mathbf{u}}$ is Eq.~(\ref{eq:Zglue_full}). 
Next, we turn to evaluate the deformation free energy. 
\begin{figure}
\begin{center}
\includegraphics[width=0.46\textwidth]{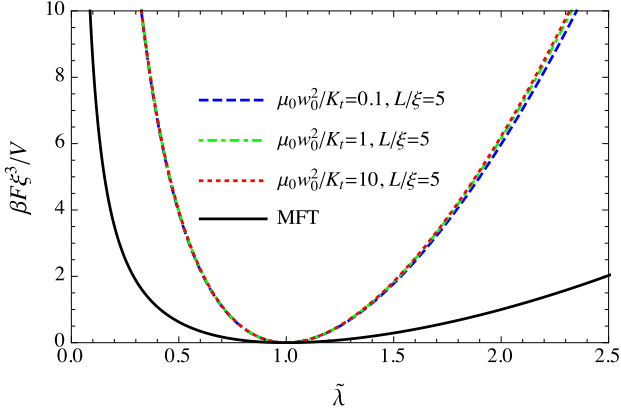}
\end{center}
\caption{Free energy behavior of a system with a soft gluing potential, $L=5\xi$ and $\mu_0=k_{{\rm B}}T/\xi^3$, for $\mu_0 w_0^2/K_t=0.1$ (blue dashed), $\mu_0 w_0^2/K_t=1$ (green dot-dashed), and $\mu_0 w_0^2/K_t=10$ (red dotted). For comparison we display the behavior predicted by classical rubber elasticity (black), with $L=\xi$ and $\mu_0=k_{{\rm B}}T$.} 
\label{fig:glue}
\end{figure}

\subsection{Free energy}

The fluctuation free energy is evaluated in Appendix~\ref{appc}, and the result is given by  
\be
\label{eq:Fu_glue}
F_{\mathbf{u}} = -k_{{\rm B}} T \ln Z = F_1 + F_2 + {\rm const},
\ee
where the ``const" refers to contributions that are independent of $\lambda$, and 
\ba
F_1 &\equiv& 
\frac{k_{{\rm B}}TS}{4\pi} \!\!
\int_0^{\pi\xi^{-1}} \!\!\!\!\!\!\!\! dQ\,Q
\ln(1-e^{-2\lambda^{3/2}QL})
\nonumber\\
&&- 
\frac{3\pi k_{{\rm B}}TS}{16\xi^2}\ln\lambda
+ \frac{\pi k_{{\rm B}}TV}{24\xi^3} f_{{\rm bulk}},
\label{eq:F1}
\\
F_2 &\equiv& 
\frac{k_{{\rm B}}TS}{4\pi} \!\!
\int_0^{\pi\xi^{-1}} \!\!\!\!\!\!\!\! dQ\,Q 
\, \ln \big[ 
\left( 1 + \alpha(\lambda) \coth Q L \right)^2
\nonumber\\
&&- 
\left( \alpha(\lambda) \, \cosech \, Q L \right)^2
\big]
\nonumber\\
&=& 
\frac{k_{{\rm B}}TS}{4\pi} \!\!
\int_0^{\pi\xi^{-1}} \!\!\!\!\!\!\!\! dQ\,Q 
\bigg\{ 2\ln(1+\alpha(\lambda)) - \ln(1-e^{-2QL})
\nonumber\\
&&+ \ln \bigg[ 1 - \left( \frac{1-\alpha(\lambda)}{1+\alpha(\lambda)} \right)^2 e^{-2QL} \bigg]
\bigg\},
\label{eq:F2}
\ea
where $\alpha(\lambda) \equiv \mu_0 w_0^2 Q/(\lambda K_t)$. 
The term $f_{{\rm bulk}}$ is given by Eq.~(\ref{eq:fbulk}). 
The contribution $F_1$ is the same as the fluctuation free energy for an elastomer with rigid pinning BC (cf. Eqs.~(\ref{eq:Fu_comp}) and (\ref{eq:Fusb}) of Sec.~\ref{sec:rigidpinning}), whilst $F_2$ is the extra contribution that arises from the finite strength of the gluing potential. The first term of $F_2$ can be interpreted as a surface tension term, whilst the second and third terms describe the interaction between the confining surfaces. 
As we expect, the last term in the formula for $F_2$ has a similar form to the thermal Casimir free energy for a slab with soft boundary conditions and no region exterior to the slab~\cite{dean_softBC,schmidt,cavalcanti,bajnok,romeo}. 

As in Eq.~(\ref{eq:deform_free_energy}) of Sec.~\ref{sec:rigidpinning}, the full deformation energy is given by 
\be
F(\widetilde{\lambda}) = F_{{\rm{full}}}(\widetilde{\lambda} \lambda_0) - F_{{\rm{full}}}(\lambda_0),
\ee
where $F_{{\rm{full}}} \equiv F_0 + F_1 + F_2$. 
For $L=5\xi$, we find, by numerically solving for $\lambda_0$ in the stationarity condition Eq.~(\ref{eq:F_stationarity}), viz., $(\partial F_{{\rm full}}(\lambda)/\partial \lambda)|_{\lambda=\lambda_0} = 0$, that $\lambda_0 \approx 1.05414$ for $\mu_0 w_0^2/K_t=0.1$, $\lambda_0 \approx 1.07503$ for $\mu_0 w_0^2/K_t=1$, and $\lambda_0 \approx 1.08807$ for $\mu_0 w_0^2/K_t=10$. As we do not know the actual value of the adhesion strength $K_t$, we have tried a range of values from small to large~\cite{haddadan1,haddadan2}. The corresponding deformation free energy behavior is displayed in Fig.~\ref{fig:glue}. The deformation free energy is larger for smaller gluing strengths $K_t$, because more fluctuation modes can be excited, and each mode contributes thermal energy to the overall free energy.

\section{Elastic Pseudo-Casimir stress}

We now turn to explore the fluctuation-induced, or pseudo-Casimir, stresses that lead the system to spontaneously relax to the true ground state, in particular comparing the attractive component of such stresses with the thermal Casimir stresses of non-elastomeric systems. To consider fluctuation stresses we consider deformations defined relative to the mean-field ground state (i.e., before the system has spontaneously relaxed). As is typical in studies of the Casimir effect, we will focus on the surface free energy contribution~\cite{krech}. We shall look at the effects of the rigid pinning BC and the soft gluing potential. 
\begin{figure}
\begin{center}
\includegraphics[width=0.46\textwidth]{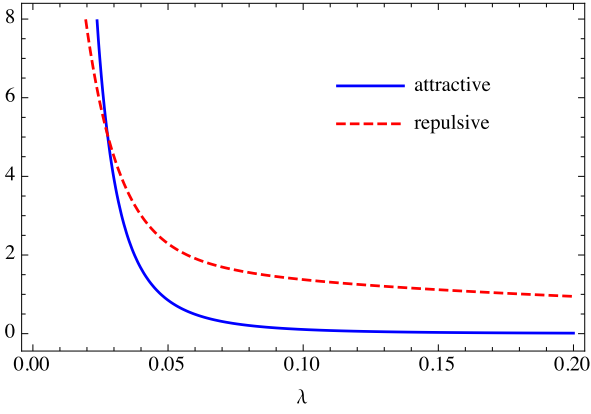}
\end{center}
\caption{Comparison (in dimensionless units) of the \emph{magnitudes} of the attractive, pseudo-Casimir term, i.e., $\zeta_{{\rm R}}(3)\xi^2/16\pi\lambda^3L^2$ (blue solid), and the repulsive and finite-size contributions to the surface free energy part of Eq.~(\ref{eq:Fusb}), i.e., $-g(\lambda,L)\xi^2/4\pi - 3\pi \ln \lambda/16$ (red dashed), for the case $L/\xi=15$. For $\lambda=0.07$ (corresponding to a measured thickness of $L'=1.05 \xi$, which is larger than the typical localization length), the magnitude of the former contribution is $18.3 \%$ of the latter contribution.} 
\label{fig:compete}
\end{figure}
First we consider the case of rigid pinning BC, and we refer to the discussion in Sec.~\ref{sec:rigidpinning}. In Eq.~(\ref{eq:Fusb}), the first term is attractive and reminiscent of a Casimir effect:
\be
F_{{\rm c}} \equiv 
-\frac{k_{{\rm B}}T S \zeta_{\rm{R}}(3)}{16\pi\lambda^3 L^2} = 
-\frac{k_{{\rm B}}T S L \zeta_{\rm{R}}(3)}{16\pi (L')^3}, 
\ee
where $L' = \lambda L$ is the thickness of the deformed elastomer.
The pseudo-Casimir contribution $F_{{\rm c}}$ is thus attractive, 
and decays as the inverse \emph{cube} of the current separation between the substrates. 
Besides this attractive contribution to the surface free energy, Eq.~(\ref{eq:Fusb}) also contains a surface term which is
proportional to $-\ln\lambda$, and thus repulsive for compressions, as well as a finite-size correction (proportional to $-g(\lambda, L)$) to the pseudo-Casimir term. These latter two contributions to the surface free energy compete with the attractive pseudo-Casimir contribution, as we see from Fig.~\ref{fig:compete}. 
Our problem is thus distinct from the pseudo-Casimir physics of confined \emph{non-elastomeric} systems such as a nematic liquid crystal confined between two flat plates with strong homeotropic anchoring at the surface of each plate. There the director fluctuation free energy decays inversely as the \emph{square} of the separation between the plates, and does not involve any additional repulsive terms originating from internal stresses of the intervening medium~\cite{ajdari}.

We determine the pseudo-Casimir stress $\sigma_{{\rm c}}$ from the formula 
\be
\sigma_{{\rm c}} = -\frac{\lambda}{S}\frac{\partial F_{{\rm c}}(\lambda)}{L\partial\lambda} = -\frac{3 k_{{\rm B}}T \zeta_{\rm{R}}(3)}{16\pi\lambda^3 L^3} = 
-\frac{3 k_{{\rm B}}T \zeta_{\rm{R}}(3)}{16\pi (L')^3}, 
\ee
which is attractive and decays as $(L')^{-3}$. The right-hand side of the first equality contains a prefactor $\lambda$ to account for the change in the cross-sectional area after deformation. This prefactor is necessary to define the \emph{true} stress (as opposed to \emph{nominal} stress, which decays as $(L')^{-4}$). 
The distinction between true and nominal stress only arises because we are dealing with an incompressible elastomer~\cite{treloar}, while in the studies of the Casimir effect in non-elastomeric systems~\cite{bordag,mostepanenko,ajdari,ziherl,jure}, the Casimir stress computed corresponds to the nominal stress. Although the pseudo-Casimir stress in a confined elastomer film decays with the same power law as that in non-elastomeric systems (e.g., confined nematic and electromagnetic fluctuations), the mechanisms giving rise to the same power law are qualitatively distinct. 

Next, we consider the correction to $F_c$ and $\sigma_c$ that come from a soft gluing potential (cf. Sec.~\ref{sec:glue}). To enable a formal comparison with the more studied case of the pseudo-Casimir effect emerging in confined nematic liquid crystals~\cite{ajdari}, consider the last term of Eq.~(\ref{eq:F2}) with the upper bound in the integral set to infinity; let us call this $\delta F_c$: 
\be
\delta F_c = \frac{k_{{\rm B}}TS}{4\pi L^2} \!\!
\int_0^{\infty} \!\!\!\! du\,u 
\ln \bigg[ 1 - \left( \frac{1-x u}{1+x u} \right)^2 e^{-2u} \bigg],
\ee
where $u \equiv QL$ and $x \equiv \mu_0 w_0^2/(\lambda K_t L)$. The problem is analytically tractable~\cite{podgornik-solvent} for sufficiently large $K_t$, where $x$ is small, and we can expand $\delta F_c$ in powers of $x$, obtaining:
\ba
\delta F_c &\approx& 
-\frac{k_{{\rm B}}TS \zeta_{\rm{R}}(3)}{16\pi L^2} 
+ \frac{k_{{\rm B}}TS \zeta_{\rm{R}}(3) \mu_0 w_0^2}{4\pi \lambda K_t L^3} 
\nonumber\\
&&- \frac{3 k_{{\rm B}}TS \zeta_{\rm{R}}(3) (\mu_0 w_0^2)^2}{4\pi \lambda^2 K_t^2 L^4}
\ea
The first term can be disregarded as it is independent of $\lambda$. The corresponding true stress is 
\ba
\delta \sigma_c &=& 
\frac{k_{{\rm B}}T \zeta_{\rm{R}}(3) \mu_0 w_0^2}{4\pi \lambda K_t L^4} 
- \frac{3 k_{{\rm B}}T \zeta_{\rm{R}}(3) (\mu_0 w_0^2)^2}{2\pi \lambda^2 K_t^2 L^5}
\nonumber\\
&=&
\frac{k_{{\rm B}}T \zeta_{\rm{R}}(3) \mu_0 w_0^2}{4\pi K_t L^3 L'} 
- \frac{3 k_{{\rm B}}T \zeta_{\rm{R}}(3) (\mu_0 w_0^2)^2}{2\pi K_t^2 L^3 (L')^2}
\ea
The leading term has a positive sign, indicating that the soft gluing potential leads to a less attractive pseudo-Casimir stress, and decays as $(L')^{-1}$. The pseudo-Casimir stress for a system with a soft gluing potential is obtained by adding $\delta \sigma_c$ to $\sigma_c$. Again, the contribution of the pseudo-Casimir stress correction is offset by the finite size correction as well as the first term of Eq.~(\ref{eq:F2}). 

\section{Summary and conclusion}

In this study we have examined the important contribution of thermal fluctuations of the elastic displacement to the elasticity of thin confined elastomer films of an axisymmetric shape. We have found that there can be significant departures (both quantitative and qualitative) from the prediction of classical rubber elasticity theory when elastic fluctuations are included. Furthermore, we have also addressed the impact of different types of boundary conditions on a particular (solenoidal) mode of elastic fluctuation on the elastic deformation free energy, focussing on the effect of (i)~a ``rigid pinning" boundary condition and (ii)~a pair of ``gluing" boundary potentials (which can be regarded as the elastic analogue of hydrodynamic slippage). We found that the deformation free energy is lower in case~(i) than in case~(ii).  
In addition, we have also explored the formal similarities of the attractive component of the elastic fluctuation-induced (pseudo-Casimir) stress with the thermal Casimir stress. The attractive component for the rigid pinning BC, as well as the leading order term corresponding to the gluing potential problem, has the same $(L')^{-3}$ decay, where $L'$ is the inter-surface separation. On the other hand, there are also corrections to the attractive part of the fluctuation stress for the gluing potential problem (which stem from the finiteness of the gluing strength), and the leading correction term decays as $(L')^{-1}$. 

Our investigation into the thermal elastic fluctuation effects between bounding surfaces in the context of confined elastomers now opens up the venue of analyzing the effective fluctuation-induced interactions between rigid inclusions in the elastomer network. Similar fluctuation mediated interactions have been introduced between e.g. protein inclusions in a background of thermal fluctuations of the lipid membrane, that exist as long as the rigidity of the inclusion differs from that of the ambient membrane \cite{li-kardar}. Another possible venue is to investigate the analogue of the \emph{critical} Casimir effect (i.e., the Casimir effect in a system characterized by an order parameter, for example a thin superfluid film, where the effect is generated by long-range fluctuations of the order parameter when the system is near-critical, such that the fluctuations approach ``masslessness"~\cite{critical1,critical2}) in elastomeric systems, for example cross-linked polymer blends under confinement. 
Our approach can be extended to study the effects of disorder introduced by random chemical cross-linking (disorder effects have indeed received a lot of attention recently in other examples of fluctuation-induced forces; see Refs. \cite{rudiali,disorder-PRL,jcp2010,pre2011,epje2012,jcp2012,disorder-book,karimi1,karimi2,li-kardar,bing2015} and references therein), and also to investigate the behavior of nematic elastomers~\cite{wt,skacej,lu-phenom,lu-IGNE} in confined environments. 

\section{Acknowledgment}
BSL thanks Institute for Research in Fundamental Sciences (IPM), Tehran, for a memorable stay in May 2015, where the present work was conceived. He also thanks L. Athanasopoulou for constructive discussions. BSL and RP would like to acknowledge the  financial support of the Agency for research and development of Slovenia under Grants No. N1-0019 and P1-0055. AN acknowledges partial support from the Royal Society, the Royal Academy of Engineering, and the British Academy (UK). 

\appendix

\section{Derivation of the fluctuation Hamiltonian, Eq.~(\ref{eq:H_u})}
\label{appa}
Here we fill in the steps leading from Eq.~(\ref{eq:Z}) to (\ref{eq:H_u}). Using Eq.~(\ref{eq:H_el}), we can express Eq.~(\ref{eq:Z}) as 
\ba
Z &\!=\!& \int\!\mathcal{D}\mathbf{u} \, 
\delta(\nabla\!\cdot\!\mathbf{u}) 
\exp\bigg\{ \!\!
-\frac{\beta \mu_0}{2} \!\int_0^L\!\!\!\! dz \!\! \int\!\! d^2x_\perp 
\bigg[ 
\Lambda_{ia} \Lambda_{ia} 
\nonumber\\
&&\quad
+ 2 \Lambda_{ia} \Lambda_{ib} \frac{\partial u_b}{\partial x_a} 
+ \Lambda_{ib} \Lambda_{ic} \frac{\partial u_b}{\partial x_a} \frac{\partial u_c}{\partial x_a} 
\bigg]
\bigg\}
\ea
Let us define a matrix $g_{ab} \equiv \Lambda_{ia}\Lambda_{ib}$ and a vector $u_a' \equiv g_{ab} u_b$. In terms of the new variable and using $\delta(\nabla\cdot\mathbf{u}) = \det \underline{\underline{g}} \, \delta(\partial_a g_{ab} u_b)$, 
 we can express $Z$ as 
\ba
Z &\!\!=\!\!& \int\!\mathcal{D}\mathbf{u}'  \mathcal{J} \, 
\delta(\nabla\!\cdot\!\mathbf{u}') \det \underline{\underline{g}} 
\nonumber\\
&&\times 
\exp\bigg\{ \!\!
-\frac{\beta \mu_0}{2} \!\int_0^L\!\!\!\! dz \!\! \int\!\! d^2x_\perp 
\bigg[ 
\Lambda_{ia} \Lambda_{ia} 
+ 2 \Lambda_{ia} \Lambda_{ib} g_{bc}^{-1} \frac{\partial u_c'}{\partial x_a} 
\nonumber\\
&&\quad
+ \Lambda_{ib} \Lambda_{ic} g_{bb'}^{-1} g_{cc'}^{-1} 
\frac{\partial u_{b'}}{\partial x_a} \frac{\partial u_{c'}}{\partial x_a} 
\bigg]
\bigg\}
\label{eq:A2}
\ea
Here $\mathcal{J} \equiv ||\delta \mathbf{u} / \delta \mathbf{u}'||$ is the functional Jacobian for the transformation of field variables $\mathbf{u}$ to $\mathbf{u}'$.  
As $\Lambda_{ia}\Lambda_{ib}g_{bc}^{-1} = g_{ab} g_{bc}^{-1} = \delta_{ac}$, the second term in the exponent is proportional to $\nabla\!\cdot\!\mathbf{u}'$. The Dirac delta-function $\delta(\nabla\!\cdot\!\mathbf{u}')$ is only non-zero for configurations for which $\nabla\!\cdot\!\mathbf{u}'=0$, implying that we can set the second term of the exponent to zero. Next we make a change of variables from $\mathbf{u}'$ back to $\mathbf{u}$. Equation~(\ref{eq:A2}) then becomes 
\be
Z = \int\!\mathcal{D}\mathbf{u} \, 
\delta(\nabla\!\cdot\!\mathbf{u}) e^{-\beta( \frac{\mu_0 V}{2} \Tr(\underline{\underline{\Lambda}}^{{\rm T}}\cdot\underline{\underline{\Lambda}}) + H_{\mathbf{u}})},
\ee
where $H_{\mathbf{u}}$ is given by Eq.~(\ref{eq:H_u}). 

\section{Derivation of the fluctuation free energy, Eq.~(\ref{eq:Fu_discrete})}
\label{appb}
The fluctuation Hamiltonian after the constraint of local incompressibility has been applied, is given by Eq.~(\ref{eq:Hfluctcon}), viz.,
\ba
H_{\mathbf{u}}[\{\psi_n, \phi_n\}] &\!\!=\!\!& \frac{\mu_0}{2\lambda}\sum_{n=1}^{\infty} \int\!\! \frac{d^2\Qv}{(2\pi)^2} 
\Big\{
\Big( Q^2 + \left( \frac{n\pi}{L} \right)^2 \Big) 
\nonumber\\
&&\quad\times 
\left[ (\psi_n^{{\rm re}}(\Qv))^2 + (\psi_n^{{\rm im}}(\Qv))^2 \right] 
\nonumber\\
&&+\Big( \lambda^3 Q^2 + \Big(\frac{n\pi}{L}\Big)^2 \Big)\Big( 1 + \Big( \frac{n\pi}{QL} \Big)^2 \Big) 
\nonumber\\
&&\quad\times 
\left[ (\phi_n^{{\rm re}}(\Qv))^2 + (\phi_n^{{\rm im}}(\Qv))^2 \right] 
\Big\}
\ea
The partition function $Z_{\mathbf{u}}$ with the incompressibility constraint and Dirichlet BC taken into account can consequently be expressed in Fourier space as 
\ba
Z_{\mathbf{u}} &=& 
\prod_{\overset{\{ \Qv > \mathbf{0} \}}{\{ n \in Z^+ \}}} \!\!\!
\!\int\! d\phi_{n}^{{\rm re}}(\Qv) d\phi_{n}^{{\rm im}}(\Qv)
\!\int\! d\psi_{n}^{{\rm re}}(\Qv) d\psi_{n}^{{\rm im}}(\Qv)
\nonumber\\
&&\quad\times 
e^{-\beta H_{\mathbf{u}}[\{ \phi_n^{{\rm re}}, \phi_n^{{\rm im}}, \psi_n^{{\rm re}}, \psi_n^{{\rm im}} \}]}
\label{eq:B2}
\ea
The Hamiltonian is Gaussian in the fluctuations, and thus the functional integrals over the fluctuating fields can be straightforwardly performed, yielding
\ba
Z_{\mathbf{u}}
&\!\!=\!\!& \!\!\!\! 
\prod_{\overset{\{ \Qv > \mathbf{0} \}}{\{ n \in Z^+ \}}}    
\!\!\!\! 
\left[ \frac{2\pi k_{{\rm B}}T \lambda S}{\mu_0\big( Q^2 + \big( \frac{n\pi}{L} \big)^2 \big)} \right] 
\nonumber\\
&&\times
\bigg[ \frac{2\pi k_{{\rm B}}T \lambda S}{\mu_0 \big( \lambda^3 Q^2 + \big( \frac{n\pi}{L} \big)^2 \big) \big( 1 + \big( \frac{n\pi}{QL} \big)^2 \big)} \bigg]
\nonumber\\
&\!\!=\!\!& e^{A -\beta F_{\mathbf{u}}(\lambda)}, 
\label{eq:B3}
\ea
where $Z^+$ refers to the set of positive integers, $\{ \Qv > \mathbf{0} \}$ refers to the set of all positive wave-vectors (``positivity" being defined with reference to a straight line that divides the two-dimensional lattice of points $(Q_x, Q_y)$ into two halves; e.g., if we denote the normal vector to such a line by $\mathbf{n}$, then a wave-vector is positive if it satisfies $\Qv \cdot \mathbf{n} > 0$), $A$ is a constant defined by
\be
A \!\equiv\! \sum_{n=1}^{\infty} \sum_{\{ \Qv > \mathbf{0} \}} \!\!\! 
\Big\{ \!  
\ln\Big[ \frac{4\pi^2S^2(k_{{\rm B}}T)^2 Q^2}{\mu_0^2} \Big] 
- 2\ln \Big[ Q^2 + \Big( \frac{n\pi}{L} \Big)^2 \Big]
\Big\} 
\ee
and $F_{{\mathbf{u}}}$ is given by 
\be
\label{eq:Fsum}
F_{{\mathbf{u}}} \equiv k_{{\rm B}}T \sum_{n=1}^{\infty} \sum_{\{ \Qv > \mathbf{0} \}} \ln \Big[ \lambda Q^2 + \frac{1}{\lambda^2} \Big( \frac{n\pi}{L} \Big)^2 \Big]
\ee
In the above, $S$ is the cross-sectional area of the surface of the elastomer film that is co-planar with the confining substrates, in the state prior to external deformation. The functional integral runs over all independent fluctuating field degrees of freedom. As the real and imaginary components of the (complex) fluctuating fields $\phi_\Qv$ and $\psi_\Qv$ obey the relations $\phi_{-\Qv}^{{\rm re}} = \phi_{\Qv}^{{\rm re}}$, $\phi_{-\Qv}^{{\rm im}} = -\phi_{\Qv}^{{\rm im}}$ (and similar ones for $\psi_\Qv$, these relations being required by the reality of the fluctuating fields in real space), the modes with positive and negative wave-vectors $\Qv$ are not really independent of each other, and thus the functional integral product runs only over the positive wave-number contributions. 
By making the continuum limit $\sum_{\{\Qv\}} = S\int d^2\Qv/(2\pi)^2$ where the wave-vector sum now runs over all wave-vectors, we have 
\be
F_{{\mathbf{u}}} = 
\frac{k_{{\rm B}}T S}{2} \sum_{n=1}^{\infty} \! \int \!\! \frac{d^2\Qv}{(2\pi)^2} \ln \Big[ \lambda Q^2 + \frac{1}{\lambda^2} \Big( \frac{n\pi}{L} \Big)^2 \Big],
\label{eq:B7}
\ee
which is Eq.~(\ref{eq:Fu_discrete}). 

\section{Derivation of the fluctuation free energy for a ``glued" elastomer, Eq.~(\ref{eq:Fu_glue})}
\label{appc}
In this Section, we provide the calculational steps to derive the fluctuation free energy in Eq.~(\ref{eq:Fu_glue}) for an elastomer ``glued" to the substrates. 
The partition function $Z_{\mathbf{u}}$ corresponding to Eq.~(\ref{eq:Hglue}) is given by Eq.~(\ref{eq:Z_glue}). 
Similar to the step from Eq.~(\ref{eq:B2}) to (\ref{eq:B3}), the evaluation of Eq.~(\ref{eq:Z1_glue}) involves functionally integrating over the sets of fluctuating fields $\{\phi_n^{{\rm re}} \}$ and $\{ \phi_n^{{\rm im}} \}$ which are Gaussian in form (as we can see from Eq.~(\ref{eq:H1_glue})). The functional integration thus yields 
\be
Z_1  = \!\!\!\!
\prod_{\overset{\{ \Qv > \mathbf{0} \}}{\{ n \in Z^+ \}}}    
\!\!\!\! \left[ \frac{2\pi k_{{\rm B}}T S}
{\mu_0\big( \lambda^2 Q^2 + \frac{1}{\lambda}\big( \frac{n\pi}{L} \big)^2 \big)
\big( 1 + \big( \frac{n\pi}{QL} \big)^2 \big)} \right]
\label{eq:Z1_gluey}
\ee
To evaluate $Z_\psi$ (cf. Eq.~(\ref{eq:Zpsi_glue})), we have to functionally integrate over six sets of fluctuating fields, viz., $\{\psi_n^{{\rm re}} \}$, $\{ \psi_n^{{\rm im}} \}$, $\{ X_\Qv^{{\rm re}} \}$, $\{ X_\Qv^{{\rm im}} \}$, $\{ Y_\Qv^{{\rm re}} \}$, and $\{ Y_\Qv^{{\rm im}} \}$. Again, as we see from Eq.~(\ref{eq:Hpsi_glue}), these fields are Gaussian in form. Making use of the formula 
\be
\int_{-\infty}^\infty \!\!\!\! dX \! \int_{-\infty}^\infty \!\!\!\! dY \, e^{-\frac{a}{2}(X^2+Y^2)+b XY} = \frac{2\pi}{\sqrt{a^2-b^2}}, 
\ee
with the identification $a = \frac{\mu_0 Q}{\lambda}\coth Q L + \frac{K_t}{w_0^2}$ and $b = \frac{\mu_0 Q}{\lambda} ( \cosech \, Q L )$, and $H_\psi$ from Eq.~(\ref{eq:Hpsi_glue}), we have 
\ba
Z_{\psi} &=\!\!\!\!& 
\prod_{\overset{\{ \Qv > \mathbf{0} \}}{\{ n \in Z^+ \}}} \!\!\!
\!\int\! d\psi_{n}^{{\rm re}}(\Qv) d\psi_{n}^{{\rm im}}(\Qv)
\!\int\! dX_{\Qv}^{{\rm re}} dX_{\Qv}^{{\rm im}}
\!\int\! dY_{\Qv}^{{\rm re}} dY_{\Qv}^{{\rm im}}
\nonumber\\
&&\quad\times 
e^{-\beta H_{\mathbf{u}}[\{ \psi_n^{{\rm re}}, \psi_n^{{\rm im}}, X_\Qv^{{\rm re}}, X_\Qv^{{\rm im}}, Y_\Qv^{{\rm re}}, Y_\Qv^{{\rm im}} \}]}
\nonumber\\
&=\!\!\!\!& 
\prod_{\overset{\{ \Qv > \mathbf{0} \}}{\{ n \in Z^+ \}}} \!\!\!
\frac{4\pi \lambda k_{{\rm B}} T S}{L \mu_0(Q^2+(\frac{n \pi}{L})^2)} 
\nonumber
\\
&&\times \frac{4\pi^2 (k_{{\rm B}} T)^2 S}{(\frac{\mu_0 Q}{\lambda} \coth Q L + \frac{K_t}{w_0^2})^2 - (\frac{\mu_0 Q}{\lambda} \cosech \, Q L)^2},
\label{eq:Zpsi_gluey}
\ea
where the first factor in the right-hand side of the second equality comes from a functional integration over $\psi_{n}^{{\rm re}}(\Qv)$ and $\psi_{n}^{{\rm im}}(\Qv)$, and the second factor comes from a functional integration over $X_{\Qv}^{{\rm re}}$, $X_{\Qv}^{{\rm im}}$, $Y_{\Qv}^{{\rm re}}$, and $Y_{\Qv}^{{\rm im}}$. 

Using Eqs.~(\ref{eq:Z1_gluey}) and (\ref{eq:Zpsi_gluey}), we find that the partition function $Z_{\mathbf{u}}$ in Eq.~(\ref{eq:Z_glue}) is given by 
\ba
\label{eq:Zglue_full}
&&Z_{\mathbf{u}} 
\\
&=& 
\prod_{\overset{\{ \Qv > \mathbf{0} \}}{\{ n \in Z^+ \}}} \!\!\!
\frac{32\pi^4(k_{{\rm B}} T)^4 w_0^4 S^3}
{L \mu_0^2 K_t^2 
Q^2 \big( 1 + \big( \frac{n\pi}{QL} \big)^2 \big)^2 
\big[ \lambda Q^2 + \frac{1}{\lambda^2} \big( \frac{n\pi}{L} \big)^2 \big]}
\nonumber\\
&&\times 
\frac{1}{(1+\alpha(\lambda) \coth QL)^2 - (\alpha(\lambda) \, \cosech \, QL)^2},
\nonumber
\ea
where $\alpha(\lambda) \equiv \mu_0 w_0^2 Q/(\lambda K_t)$. 
The fluctuation free energy is given by $F_{\mathbf{u}} = -k_{{\rm B}}T \ln Z_{\mathbf{u}}$, i.e., 
\ba
\label{eq:C5}
&&F_{\mathbf{u}} 
\\
&=& D +
k_{{\rm B}} T \!\!\! \sum_{\{\Qv > \mathbf{0}\}} 
\Big\{ 
\sum_{n=1}^{\infty} 
\ln \Big[ \lambda Q^2 + \frac{1}{\lambda^2} \left( \frac{n\pi}{L} \right)^2 \Big]
\nonumber\\
&&+
\ln \big[ (1+\alpha(\lambda) \coth QL)^2 - (\alpha(\lambda) \, \cosech \, QL)^2 \big]
\Big\},
\nonumber
\ea
where $D$ is independent of $\lambda$, given by 
\ba
D &\equiv& -k_{{\rm B}} T \!\!\! \sum_{\{\Qv > \mathbf{0}\}} 
\Big\{
\ln \frac{32\pi^4 (k_{{\rm B}} T)^4 w_0^4 S^3}{L \mu_0^2 K_t^2}
\nonumber\\
&&\quad
- \sum_{n=1}^{\infty} \ln Q^2 \Big[ 1 + \Big( \frac{n\pi}{QL} \Big)^2 \Big]^2
\Big\}
\ea
The second term of Eq.~(\ref{eq:C5}) is identical to the term in Eq.~(\ref{eq:Fsum}), which we can write as $F_1$, where 
\be
F_1 \equiv \frac{k_{{\rm B}}T S}{2} \! 
\sum_{n=1}^{\infty} \! \int \!\! \frac{d^2\Qv}{(2\pi)^2}
\ln \Big( \lambda Q^2 + \frac{1}{\lambda^2}\Big( \frac{n\pi}{L} \Big)^2 \Big)
\ee
Making use of Eqs.~(\ref{eq:finiteintegral}) and (\ref{eq:Fusb}) allows us to rewrite $F_1$ as Eq.~(\ref{eq:F1}). 

We can rewrite the third term of Eq.~(\ref{eq:C5}) as $F_2$, where
\ba
F_2 
&\equiv\!\!& 
k_{{\rm B}} T \!\!\! \sum_{\{\Qv > \mathbf{0}\}} \!\!\!
\ln \big[ (1+\alpha(\lambda) \coth QL)^2 
\nonumber\\
&&\qquad\qquad
- (\alpha(\lambda) \, \cosech \, QL)^2 \big]
\nonumber\\
&=&
\frac{k_{{\rm B}} T S}{2} \!\! \int \! \frac{d^2\Qv}{(2\pi)^2} 
\ln \big[ (1+\alpha(\lambda) \coth QL)^2 
\nonumber\\
&&\qquad\qquad
- (\alpha(\lambda) \, \cosech \, QL)^2 \big].
\ea
In the second step we have made the continuum limit, and we thus arrive at Eq.~(\ref{eq:F2}). Summing up the contributions $D$, $F_1$ and $F_2$ gives us the fluctuation free energy, Eq.~(\ref{eq:Fu_glue}).


\begin{thebibliography}{99}
\bibitem{footnote:thicknesses}
These values for $L$ are consistent with the thicknesses of rubber films studied, e.g. in Ref.~\cite{zhai_mckenna}, if we take the typical localization length to be of nanometer order (see, e.g., Ref.~\cite{goldbart_vulcan}). The thicknesses of ultrathin polyurethane rubber can range between 22 nm and 220 nm.
\end{thebibliography}
\end{document}